\documentstyle[aps,twocolumn,floats,epsf]{revtex} 
\begin{document}  
\draft
\title{Relations Between Quantum and Classical Spectral Determinants 
(Zeta--Functions)}                               
\author{O. Agam$^\dag$, A. V. Andreev$^{\dag \ddag}$, and B. L. 
Altshuler$^{\dag \ddag}$}  
\address{$^\dag$ NEC Research Institute, 4 Independence Way, Princeton,
 NJ 08540, USA \\
$^\ddag$ Department of Physics, Massachusetts Institute of Technology, 77  
Massachusetts Avenue, Cambridge, MA\ 02139, USA\\}
\maketitle     
\begin{abstract}

We demonstrate that beyond the universal regime
correlators of quantum spectral determinants $\Delta(\epsilon)=\det 
(\epsilon-\hat{H})$ of chaotic systems, defined through an averaging 
over a wide energy interval, are determined by the underlying
classical dynamics through the spectral determinant $1/Z(z)=\det 
(z- {\cal L})$, where $e^{-{\cal L}t}$ is the Perron-Frobenius operator.
Application of these results to the Riemann zeta function, 
allows us to conjecture new relations satisfied by 
this function. 
\end{abstract}     
\par                                         
\vspace{0.5cm} 
One of the central themes in the study of the quantal behavior of
systems which have chaotic classical counterparts, is the quest for a
correspondence between their quantal and classical
properties. Examples of such correspondences are the random matrix 
theory (RMT) \cite{Mehta91} behavior of chaotic systems over small energy 
intervals \cite{Bohigas84}, and the ``scar'' phenomenon in which wave 
functions of chaotic systems are peaked in the vicinity of short isolated 
unstable classical orbits \cite{Heller84}. 
The universal RMT behavior of chaotic systems is associated with the
long time dynamics of the chaotic system, when classical probability 
distributions have already decayed into the ergodic stationary state. 
In the field theoretic approach to disordered metals \cite{Efetov83} 
RMT emerges as the zero mode contribution associated with the ergodic state.
Scars, on the other hand, manifest individual non-universal
properties of chaotic systems associated with the short time behavior
of the classical system.

The correspondence between the quantum spectral statistics
and the classical behavior  of chaotic systems is understood,
usually, on a qualitative level. 
The purpose of this letter is to quantify this correspondence by 
formulating relations between spectral determinants 
associated with both the  quantal and the classical dynamics  of 
chaotic systems. More precisely, we will show that correlators of
spectral determinants of a given quantum system, defined through an
averaging over a large energy interval, can be expressed in terms of 
spectral determinants of the classical counterpart. 

Correlators of spectral determinants of quantum chaotic systems have 
been previously discussed in the literature \cite{Anton-Ben}. 
The results, however, were limited to the universal regime
(zero mode approximation) where RMT applies. 
Therefore, they are incapable of providing any information 
regarding individual features of the underlying 
classical dynamics. To this end it is crucial to go beyond 
the zero mode approximation, to encompass non-universal 
properties of the system. One possibility is to use periodic orbit theory. 
The non-universal properties, in this approach, are related to
short classical orbits. This approach has drawbacks which 
will be discussed below.

In this paper we will employ an alternative 
semiclassical approach based on a recently developed supersymmetric 
non-linear $\sigma$ model for chaotic systems  \cite{AASA} 
which is free of these drawbacks. 
The basic classical ingredients of this field theory
are modes of the time evolution of the underlying classical system. 
These are  the Perron--Frobenius  modes, by which a disturbance in the 
classical probability density of chaotic system relaxes into the ergodic 
distribution \cite{Ruelle}. 
They are analogous to the diffusion modes in the case 
of disordered systems \cite{AAA}. 
For time $t>0$, the $\mu$-th Perron--Frobenius 
mode decays as $e^{-\gamma_\mu t}$ where $\Re \gamma_\mu \geq 0$. 
$\gamma_\mu$ are known as the Ruelle resonances. 
The dynamical zeta function $1/Z(z)$ is the 
classical spectral determinant defined as
\begin{equation}
1/Z(z)= \prod_\mu A_\mu (z-\gamma_\mu), \label{Cspectral}
\end{equation}
where $A_\mu$ are regularization factors introduced to make
the product converge. This function, for systems which are
ergodic on the energy shell, has a simple zero at the origin, 
$1/Z(0)= 0$. This zero is associated with
the ergodic state to which probability densities evolve, and
manifests the conservation of probability. Thus $Z(z)$, in the 
limit $z \to 0$, takes the form
\begin{equation}
\lim_{z \to 0} Z(z) = -\frac{U}{\pi z}+ V \label{Zlimit}
\end{equation}
where $U$ and $V$ are real constants. 

The quantum spectral determinant which we consider here is
associated with the spectrum of the Hamiltonian $\hat{H}$. It 
is a real function of the energy $\epsilon$ whose zeros coincide 
with the  eigenvalues  $\epsilon_n$ of $\hat{H}$,
\begin{equation}
\Delta(\epsilon) =\prod_n A'_n(\epsilon-\epsilon_n), 
\label{Qspectral}
\end{equation}
where $A'_n$ are regularization factors.

The correlators of the quantum spectral determinants 
will be defined with respect to the energy averaging:
\begin{equation}
\langle \cdots \rangle = \int \frac{d \epsilon }{\sqrt{2 \pi} N}
\exp \left\{ -\frac{(\epsilon - \epsilon_0)^2}{2 N^2} \right\} ( \cdots ),
\label{EAveraging}
\end{equation} 
where $\epsilon_0$ and $N$ denote, respectively, the 
center and the width of the energy band over which averaging 
is performed. For convenience
we shall work with dimensionless quantities, i.e. all energies will be
measured in units of the mean level spacing, and $\hbar=1$.
Let  $\gamma_1$ be the first non-vanishing Perron-Frobenius eigenvalue.
We will assume that $\Re \gamma_1 \ll N \ll \epsilon_0$. Under these
conditions, results are independent of $N$, and the classical dynamics
is essentially the same over the band.

We consider here chaotic systems with no discrete symmetries
belonging to the unitary ensemble (broken time reversal symmetry). 
The formulae which will be derived and discussed in the following are:
\begin{equation}
\left\langle \frac{ \Delta (\epsilon+s) \Delta (\epsilon-s)}
{\langle \Delta^2 (\epsilon) \rangle}\right\rangle = \frac{1}{U+V} 
\Re \left\{  
Z(2is) e^{-i2\pi s} \right\}, \label{DD}
\end{equation}
\begin{eqnarray}
\left\langle \frac{\langle \Delta^2 (\epsilon) \rangle}
{ \Delta (\epsilon+\omega^+) \Delta (\epsilon-\omega^+) } \right\rangle
=\frac{2}{U+V}Z(2i\omega^+) e^{i2\pi\omega^+}, \label{1/DD}
\end{eqnarray}
where $\omega^+=\omega+i0$, and
\begin{eqnarray}
\left\langle \frac{\Delta (\epsilon+s) \Delta (\epsilon-s)}{
\Delta (\epsilon+\omega^+) \Delta (\epsilon-\omega^+)} \right\rangle
= ~~~~~~~~~~~~~~~~~~~~~~~ \label{DD/DD} \\ 
~~~~=\frac{Z(2is)Z(2i\omega^+)}{Z^2(is+i\omega^+)} e^{i2\pi (\omega^+-s)}
+ (s \to -s), \nonumber
\end{eqnarray}
which apply if $s,\omega \ll N$. When $s,\omega \ll 
\Re \gamma_1$, $Z(z)$ can be approximated by (\ref{Zlimit}) and
the RMT results \cite{Anton-Ben} are recovered.

An example  different from RMT in which these relations are {\em exact}
is the one dimensional harmonic oscillator $H=(p^2+q^2)/2$. 
This system is not chaotic, nevertheless, it is ergodic and its 
spectrum $\epsilon_n= n+1/2$ is rigid. The quantum spectral 
determinant at high energies is simply $\Delta (\epsilon)=\cos(\pi \epsilon)$, 
while the classical dynamical zeta function is 
$1/Z(z)=1-e^{2 \pi z}$ with  $U=V=1/2$.
Expanding $1/\cos (\epsilon^+) = 2 \sum_{n=0}^{\infty} 
(-1)^n e^{i\epsilon^+(2n+1)}$ and averaging over the energy it 
is straightforward to check all three relations above 
are satisfied {\em exactly}. Notice that in contrast with chaotic
systems, in this system all zeros of $1/Z(z)$ are purely imaginary.

The derivation of the spectral determinant relations for general 
chaotic systems consists of two elements: 
One is the field theoretic framework of a semiclassical 
approximation in which the classical ingredients are the Perron-Frobenius 
modes \cite{AASA}, and  the second is a method of evaluating
non-perturbative parts of correlators \cite{AA}.  

In order to sketch the derivation, we consider relation 
(\ref{DD/DD}) first. Our starting point is a formula for the ratio of
four spectral determinants. Introducing the eight component field
$\Psi^T=(\psi_R, \psi_R^*, \chi_R, \chi_R^*,\psi_A, \psi_A^*, \chi_A, 
\chi_A^*)$, where $\psi$ and $\chi$ denote commuting and anti-commuting 
variables respectively, while $R/A$ designate retarded/advanced components,
we have \cite{Efetov83},
\begin{eqnarray}
\frac{\Delta (\epsilon+s^+) \Delta (\epsilon-s^+)}{
\Delta (\epsilon+\omega^+) \Delta (\epsilon-\omega^+)} = ~~~~~~~~~~~~~~~~~~
~~~~~~\\
~~=\int {\cal D} \Psi \exp \left\{-\frac{i}{2}\int dq \Psi^{\dagger}({\bf q}) 
\Lambda \left[\epsilon-\hat{H}-F \right]  \Psi({\bf q})\right\}\nonumber
\end{eqnarray}
where ${\bf q}$ is spatial coordinate vector, $\Lambda=\mbox{diag}
(1,1,1,1,-1,-1,-1,-1)$, and $F=(\tilde{F},-\tilde{F})$ with 
$\tilde{F}=\mbox{diag}(\omega^+,\omega^+,s^+,s^+)$. 
From here we proceed along the route discussed in 
Ref. \cite{AASA}.  Namely, an energy averaging (\ref{EAveraging}) is 
performed which induces a quartic interaction among the $\Psi$ fields.
Then, this interaction is decoupled by means of the Hubbard--Stratonovich 
transformation, and the integration over $\Psi$ is performed. 
Finally, a steepest descent approximation, in which the width 
of the band $N$ plays the role of the large parameter, is
used to integrate out the massive modes. The result is of the form of
an integral over the remaining Goldston (massless) modes,
\begin{eqnarray}
\left\langle \frac{\Delta (\epsilon+s^+) \Delta (\epsilon-s^+)}{
\Delta (\epsilon+\omega^+) \Delta (\epsilon-\omega^+)}\right\rangle =
\int {\cal D} T \exp \left( S_{eff} \right) \label{Functional-Int} \\
S_{eff}=\frac{\pi}{2} \int d x_\parallel \mbox{STr} \left[
\left( iF+T^{-1} {\cal L}T \right) 
Q \right],
\end{eqnarray}
where $Q =T^{-1} \Lambda T$, and $T=T({\bf x}_\parallel)$ belongs 
to the coset space $U(1,1/2)/[U(1/1)\otimes U(1/1)]$ \cite{Zirnbauer86}. 
Here ${\bf x}_\parallel$ denotes a vector of phase space variables 
which lies on the energy shell $\epsilon_0=H({\bf x})$, and
normalized such that $\int dx_\parallel = 1$.
${\cal L}$ is the generator of time evolution of the classical
dynamics defined by the Poisson brackets,
\begin{equation}
{\cal L} \ \cdot = \{ H, \ \cdot \ \}.
\end{equation}
The eigenvalues of ${\cal L}$ arising from the  regularization
of the functional integral (\ref{Functional-Int}) 
constitute the Perron-Frobenius spectrum $\{ \gamma_\mu \}$.

The large frequency assymptotics of the above integral can be calculated
by stationary phase following Ref. \cite{AA}. Perturbation theory 
corresponds to the integration of small fluctuations of $T$ 
around unity, so that $T=(1-iP)^{-1}$ where $P$ is small, and
\begin{eqnarray}
Q=\Lambda (1+iP)(1-iP)^{-1}, ~~~ P=\left( \begin{array}{cc} 0 & B \\
\bar{B} & 0 \end{array} \right).
\end{eqnarray}
However,  $Q=\Lambda$ is not the only stationary point on the coset space
$U(1,1/2)/[U(1/1)\otimes U(1/1)]$. For the unitary ensemble 
there exists a second stationary point which is $Q=k\Lambda$ 
where $k=\mbox{diag}(1,1,-1,-1,1,1,-1,-1)$ \cite{AA}. In our case it amounts 
merely to a change of sign in the Fermion block, namely 
to the substitution of $s \to -s$ in the final results obtained from the
first stationary point. 
It is therefore sufficient to consider the fluctuations around
$Q=\Lambda$. Expanding the action to the leading quadratic order 
in $B$ and $\bar{B}$ we have
\begin{eqnarray}
S_{eff} \simeq i2\pi (\omega^+-s) + ~~~~~~~~~~~~~~~~~~~~~~~~~~  \\
~~+i\pi \int dx_\parallel
\mbox{Str} \left[ \tilde{F}B\bar{B}+\tilde{F}\bar{B}B
+i\bar{B}{\cal L}B\right]. \nonumber
\end{eqnarray}
Expressing $B({\bf x}_\parallel)$ and $\bar{B}({\bf x}_\parallel)$
in terms of the eigenbasis of the left $\tilde{\phi}_\nu$
and right $\phi_\mu$ Perron-Frobenius eigenfunctions, 
as $B({\bf x}_\parallel)=\sum_\mu B_\mu \phi_\mu({\bf x}_\parallel) $ and 
$\bar{B}({\bf x}_\parallel)=\sum_\nu \bar{B}_\nu \tilde{\phi}_\nu(
{\bf x}_\parallel)$, we obtain
\begin{eqnarray}   
S_{eff} \simeq 2\pi \sum_\mu \left\{ 
i(\omega^+-s)(|B_\mu^{11}|^2-|B_\mu^{33}|^2)+ \right. \nonumber \\
\left. -(\gamma_\mu-is-i\omega^+) \mbox{Str}(\bar{B}_\mu B_\mu)
 \right\}+i2\pi(\omega^+-s).
\end{eqnarray}
We remark, here, that the Perron-Frobenius eigenmodes lie outside
the Hilbert space where $B({\bf x}_\parallel)$ are defined. Nevertheless,
they form a biorthonormal set, $\int dx_\parallel
\tilde{\phi}_\nu({\bf x}_\parallel)\phi_\mu ({\bf x}_\parallel)
=\delta_{\nu,\mu}$, which spans this space \cite{AASA}. 

Integrating over $B_\mu$, and taking into account also the 
contribution of the second stationary point
$Q=k\Lambda$ we obtain (\ref{DD/DD}). 
Our perturbation theory can be justified only 
in the large frequency assymptotics $s,\omega \gg 1$, 
nevertheless, comparing it with the exact universal expression 
at the low frequency limit \cite{Anton-Ben} and with the result 
of renormalization group treatment \cite{Kravtsov94}, we conclude
that (\ref{DD/DD}) is correct for arbitrary values of $s$ and $\omega$.

Relations (\ref{DD}) and (\ref{1/DD}) can be similarly obtained 
by performing the energy averaging only in the Fermion--Fermion
or in the Boson--Boson blocks respectively. 
The normalization constants of the resulting 
expressions are determined by considering the limits $s, \omega^+ \to 0$.   

It is instructive to show how one might use periodic orbit theory 
in order to argue that the spectral determinant relations are 
satisfied. One can express both spectral determinants,
$\Delta(\epsilon)$ (in the semiclassical approximation) and $1/Z(z)$, 
in terms of the classical periodic orbits of the system.
For concreteness we consider two dimensional systems. 
The spectral determinant (\ref{Qspectral}), in the semiclassical 
approximation, can be written as \cite{Berry90}
\begin{equation}
\Delta(\epsilon)= e^{-i\pi \bar{N}(\epsilon)} \zeta_s(\epsilon),
\label{DS}
\end{equation} 
were $ \bar{N}(\epsilon)$ is the mean level staircase, and $\zeta_s(\epsilon)$ 
is the Selberg zeta function, 
\begin{equation}
\zeta_s(\epsilon) = \prod_p\prod_{k=0}^{\infty} \left( 1-
\frac{e^{iS_p(\epsilon)-i\nu_p}}{|\Lambda_p|^{1/2} \Lambda_p^k}\right).
\label{Selberg}
\end{equation} 
Here $p$ labels the primitive periodic orbits of the system
characterized by action $S_p(\epsilon)$,
Maslov phase $\nu_p$, and the eigenvalue of the monodromy matrix
$\Lambda_p$ with absolute value larger than one.

The dynamical zeta function (\ref{Cspectral}) has an exact representation
in terms of a product over periodic orbits \cite{Ruelle},
\begin{eqnarray}
1/Z(z)= \prod_p \prod_{k=0}^{\infty}\left(1-
\frac{e^{zT_p}}{|\Lambda_p| \Lambda_p^k}\right)^{k+1} 
\label{dynamical} 
\end {eqnarray}
where $T_p= \partial S_p(\epsilon)/\partial \epsilon$ is 
the period of the $p$-th orbit.

Both periodic orbit products, (\ref{Selberg}) and (\ref{dynamical}), 
suffer from convergence problems in the regions
of interest, nevertheless they imply a particular form of
regularization since they can be understood as
analytically continued from domains of the complex plain 
where the products converge. Algebraic 
manipulation with their forms should  be therefore 
understood as either performed using resummed formulae, 
or analytically continued from regions of complex energies 
($\epsilon$ and $z$) where the products converge.

To calculate $\langle \Delta(\epsilon + s)\Delta(\epsilon -s)\rangle$ 
one might try to express (\ref{Selberg}) as a Dirichlet sum over 
pseudo-orbits \cite{Berry90}, and then keep the diagonal part of the resulting
double sum in $\Delta(\epsilon + s)\Delta(\epsilon -s)$. However
such calculation yields a wrong result. The reason is that the 
pseudo--orbits in  the Dirichlet representation 
of the the Selberg zeta function are not independent.
They are related by the fact that $\Delta(\epsilon)$ for real $\epsilon$
is a real function while its divergent periodic orbit representation is not 
manifestly real. In fact, a resummation of the  divergent tail the 
Dirichlet sum reproduces the complex conjugate of the head,
and the resulting expression for the spectral determinant has a
Riemann--Siegel form in which only a finite number of orbits contribute
effectively \cite{Berry92}. Using such a formula for $\Delta(\epsilon)$
one can show that to the leading order in $1/|\Lambda_p|$ (\ref{DD}) is 
reproduced. In the case of inverse spectral determinants one should interpret 
$1/\zeta_s(E+\omega^+)$ as the usual periodic orbit product 
(\ref{Selberg}) evaluated at $\omega$ with sufficiently large imaginary part.
Applying the diagonal approximation with the above interpretation of 
the spectral determinants at the numerator and the denominator and 
using $|\Lambda_p|\gg 1$, one can obtain (\ref{1/DD}) 
and (\ref{DD/DD}).

It is worth pointing out the differences between the field theoretic
and the traditional semiclassical approaches. The starting point of
the periodic orbit theory is a semiclassical approximation
for the quantum spectral determinants given in terms of periodic orbits
(\ref{Selberg}). Energy averaging is then performed 
using the diagonal approximation. The problem is that the 
diagonal approximation is uncontrolled. Moreover it leads to wrong 
results whenever the diverging periodic orbit sums involved in the 
calculation are interpreted incorrectly. The results of this approach 
also appear to be only a leading order in the instability of the 
orbits namely when $|\Lambda_p|\gg 1$.
In the field theoretic approach the starting point is an exact quantum
mechanical expression for the spectral determinants. The energy averaging 
is performed exactly,  and only then the  semiclassical approximation 
is applied. Here the approximations at each stage of the calculation  
are controlled and large instability, $|\Lambda_p|\gg 1$, 
is not required. 
 
Our results can be used in order to reveal new mathematical relations
satisfied by the Riemann zeta function,
\begin{equation} 
\zeta(x)=\prod_p \left(1-\frac{1}{p^{x}}\right)^{-1},
\end{equation}
where the product is over all prime numbers
$p$. In view of the Riemann hypothesis, it is believed that there exists 
a chaotic Hamiltonian whose eigenenergies coincide with the non--trivial
zeros of $\zeta(x)$ on the critical line $\Re x = 1/2$ \cite{Titchmarsh}. 
Thus $\zeta(1/2-i\epsilon)$ plays the role of the Selberg zeta function
for this system. The spectral determinant (which is a real
function for real energies) is given by \cite{Titchmarsh}
\begin{equation}
\Delta_R(\epsilon) = e^{-i\pi \bar{N}(\epsilon)}\zeta(1/2-i\epsilon),
\label{DR}
\end{equation}
where $\bar{N}(\epsilon)=\frac{\epsilon}{2\pi} \left( \ln \frac{\epsilon}{2\pi}-1\right) +\frac{7}{8} + O\left(\frac{1}{\epsilon}\right)$ 
is the mean counting function, which gives the mean number of zeros below
$\epsilon$. The mean density of zeros $\bar{d}(\epsilon)= d
\bar{N}(\epsilon)/ d \epsilon \sim \ln (\epsilon)/2\pi$, thus, increases 
logarithmically with the energy. Here, for convenience, we choose 
{\em not} to normalize the energy by the mean level spacing. 

As we have shown, correlators of quantum spectral determinants are
related to the classical spectral determinant. To apply our 
results it is therefore  necessary to determine the function $1/Z_R(z)$ 
which plays the role of the  dynamical zeta function for 
the ``Riemann system''.  This can be deduced from the exact 
periodic orbit formula \cite{Cvitanovich},
\begin{equation}
1/Z(z) = \exp \left\{ - \sum_p \sum_{r=1}^{\infty}
\frac{1}{r} \frac{e^{zT_pr}}{|\det (M_p^r -I) |} \right\}, \label{Gen}
\end{equation}
where $M_p^r$ is the monodromy matrix of the $p$-th orbit repeated
$r$ times [for two dimensional systems Eq. (\ref{Gen} reduces 
to (\ref{dynamical})]. From, $T_p=\log p$ and $|\det (M_p^r -I) 
|=e^{r\ln p}$ \cite{Berry85} it follows that
\begin{equation}
Z_R(z)= \zeta(1-z), \label{Zeta-R}
\end{equation}
Thus the inverse of the dynamical zeta function
is itself the Riemann zeta function. It has a 
single simple zero at the origin, and since $\lim_{z \to 0} 
\zeta(1-z)=-1/z+C$,  $U=\pi$ and $V=C$ where 
$C=0.577215\cdots$ is Euler's constant. Assuming the existence 
of an underlying chaotic ``Riemann system'', and using 
(\ref{DR}) and (\ref{Zeta-R}) one can apply 
(\ref{DD}), (\ref{1/DD}) and  (\ref{DD/DD}), to the Riemann zeta function .

It is worth mentioning that the periodic orbit approach,  
namely using the Dirichlet representation, 
$\zeta(x)=\sum_{n=1}^{\infty} n^{-x}$, the fact that 
$\Delta_R(\epsilon)$ is real, and diagonal approximation, 
yields $\left\langle \Delta_R(\epsilon+s)\Delta_R(\epsilon-s)
\right\rangle= \Re \left\{ Z_R(2is) e^{-i2\pi \bar{d}(\epsilon_0)s}
\right\}$ which is analogous to (\ref{DD}), and exact 
to any order in $1/|\Lambda_p|$. In this respect the diagonal 
approximation for the Riemann zeta function is compatible with the 
field theoretic calculation.

Relation (\ref{1/DD}) for the Riemann zeta function can be written in 
a particularly simple form. Substituting (\ref{DR}) and (\ref{Zeta-R}) in
(\ref{1/DD}) leads to    
\begin{equation}
\left\langle \frac{1}{\zeta[\frac{1}{2}-i(\epsilon+\omega^+)]}
\frac{1}{\zeta^*[\frac{1}{2}-i(\epsilon-\omega^+)]}
\right\rangle = \zeta(1-i2\omega^+).\label{zeta}
\end{equation}

To summarize, we have shown that correlators of quantum spectral 
determinants of chaotic systems are expressed in terms of the classical
spectral determinant associated with the Perron--Frobenius spectrum.
On energy scales smaller than $\Re \gamma_1$ where $\gamma_1$ is the first
non-vanishing Perron-Frobenius eigenvalue, the universal results are
reproduced, while over larger scale the behavior is determined by 
properties of the dynamical zeta function. The results, applied
to the Riemann zeta function, enabled us to conjecture the existence
of new relations satisfied by $\zeta(s)$.
Our relations (\ref{DD}) (\ref{1/DD}) and (\ref{DD/DD}) can be
also straightforwardly generalized to higher order correlators.

We are grateful to N. Brenner, S. Fishman, A.~M.~Polyakov, D.~Ruelle, 
B. D. Simons, and Ya.~G.~Sinai, for stimulating discussions. The work 
of A.~V.~A. was supported in part by JSEP No. DAAL 03-89-0001. 
O.~A.~acknowledges the support of the Rothschild Fellowship.

\end{document}